\documentclass[twocolumn]{aastex62}
\usepackage{amsmath}
\usepackage{xspace}
\newcommand{\kep}{\emph{Kepler}\xspace}

\graphicspath{{./}{figures/}}
\shorttitle{In-situ Resonance Chains}
\shortauthors{Morrison et al.}

\begin{document}

\title{Chains of Planets in Mean Motion Resonances Arising from Oligarchic Growth}

\correspondingauthor{Sarah J. Morrison}
\email{SJMorrison@missouristate.edu}

\author[0000-0002-2432-833X]{Sarah J. Morrison}
\affil{Department of Physics, Astronomy, \& Materials Science, Missouri State University, Springfield, MO 65897}
\affil{Center for Exoplanets and Habitable Worlds}
\affil{Dept. of Astronomy \& Astrophysics
The Pennsylvania State University \\
University Park, PA 16802}

\author[0000-0001-9677-1296]{Rebekah I. Dawson}
\affil{Center for Exoplanets and Habitable Worlds}
\affil{Dept. of Astronomy \& Astrophysics
The Pennsylvania State University \\
University Park, PA 16802}

\author[0000-0003-2372-1364]{Mariah MacDonald}
\affil{Center for Exoplanets and Habitable Worlds}
\affil{Dept. of Astronomy \& Astrophysics
The Pennsylvania State University \\
University Park, PA 16802}
\nocollaboration

\begin{abstract}
Exoplanet systems with multiple planets in mean motion resonances have often been hailed as a signpost of disk driven migration. Resonant chains like Kepler-223 and Kepler-80 consist of a trio of planets with the three-body resonant angle librating and/or with a two-body resonant angle librating for each pair. Here we investigate whether close-in super-Earths and mini-Neptunes forming in situ can lock into resonant chains due to dissipation from a depleted gas disk. We simulate the giant impact phase of planet formation, including eccentricity damping from a gaseous disk, followed by subsequent dynamical evolution over tens of millions of years. In a fraction of simulated systems, we find that planets naturally lock into resonant chains. These planets achieve a chain of near-integer period ratios during the gas disk stage, experience eccentricity damping that captures them into resonance, stay in resonance as the gas disk dissipates, and avoid subsequent giant impacts, eccentricity excitation, and chaotic diffusion that would dislodge the planets from resonance. Disk conditions that enable planets to complete their formation during the gas disk stage enable those planets to achieve tight period ratios $\le 2$ and, if they happen to be near integer period ratios, lock into resonance. Using the weighting of different disk conditions deduced by MacDonald et al. (2020) and forward modeling \kep selection effects, we find that our simulations of in situ formation via oligarchic growth lead to a rate of observable trios with integer period ratios and librating resonant angles comparable to observed \kep systems.
\end{abstract}

\keywords{planets and satellites: dynamical evolution and stability --- 
planet–disk interactions --- planets and satellites: formation}

\section{Introduction} \label{sec:intro}

An enduring question about many of the several thousand known exoplanets is whether they formed where we observe them today or underwent migration from their birthplace.  The prevalence of migration (i.e., whether planets typically form where we observe them) affects our assumptions about their compositions and habitability. Even if we believe that migration likely does take place to some degree, it is unclear whether it is a vital process shaping the architectures of most planetary systems, a minor process tweaking planetary architectures, and/or a process important in some systems but not others. 

Short period planets were once assumed to be the product of migration (e.g., \citealt{lin96,rasi96,lee02}). Migration has been invoked to explain the presence and properties of super-Earths and mini-Neptunes orbiting close to their star (e.g., \citealt{ida08,coss14,izid17,izid19,carr19}) discovered in abundance by ground-based radial-velocity surveys (e.g., \citealt{howard10,mayor11}) and the \kep Mission (e.g., \citealt{how12}). However, recent simulations of planet formation have shown that planets with modest gas envelopes may be able to form close to their host stars (e.g., \citealt{chia13,lee14}), or at least assemble from transported building blocks (e.g., \citealt{hans12}). In situ formation can account for many of the observed orbital and composition properties of super-Earths and mini-Neptunes (e.g., \citealt{daws15,Lee15,daws16,mori16,macd20}). However, even if the bulk of short period super-Earths and mini-Neptunes form in situ, it has usually been assumed that at least planets in an orbital configuration known as a resonant chain must have formed elsewhere and migrated to their present location during the proto-planetary disk phase of that system's history.

Resonant chains consist of three or more planets in the same system with integer period ratios. We observe this configuration among certain moons in our solar system (most famously Io, Ganymede, and Europa) and about a dozen extra-solar systems (e.g.,\citealt{fabrycky10,gill17,macd16,mill16}). To formally be in resonance, bodies must exhibit libration of their resonant angles. We consider two types of resonant chains here. In the first type, three planets are configured in a three-body resonance with a resonant angle involving the longitude of each planet. In the second type, the inner pair of three planets and the outer pair each participates in a two-body resonance with a resonant angle involving the longitudes of the two planets and the periapse angle of one planet. We will define the various angles we consider in Section \ref{sec:sysres}.

Disk driven migration can form resonant chains (e.g., \citealt{cres06,miga16,sun17,gall16,char18}), so resonant chains have been considered a hallmark of planets that formed far from where we observe them today. Yet, in practice, MMR chains have been established in simulations not only by long distance migration (a factor of $\sim 10$ change in semi-major axis) but also short distance migration (e.g., \citealt{macd16}; a factor of $\sim 0.1$ change in semi-major axis) and via eccentricity damping only (e.g., \citealt{dong16,chok20}). Resonance widths are wider at low eccentricities (e.g., \citealt{malh20} and references therein), so eccentricity damping can expand the range of period ratios compatible with libration (e.g., \citealt{del12}).  \citet{macd18} recently showed that all three dynamical histories (long distance migration, short distance migration, and eccentricity damping only) can account for the orbital resonances observed in the Kepler-80, Kepler-223, Kepler-60, and TRAPPIST-1 planetary systems and are consistent with observed values of and constraints on their eccentricities, period ratios, libration centers, and libration amplitudes. \citet{macd18} demonstrated as a proof of concept that even systems with resonant chains may have formed at or near their observed locations. However, more work is needed to demonstrate that in situ formation is a feasible origins scenario for resonant chains. \citet{macd18}'s simulations began with fully formed planetary systems and treated the migration and eccentricity damping timescales and initial eccentricities and orbital periods as free parameters. We need to investigate whether the successful parameters are not too fine tuned or inconsistent with plausible proto-planetary disk conditions. We also need to explore whether disk conditions (e.g., \citealt{Goldreich:2014}) and perturbations from additional undetected planets (e.g., \citealt{pan17}) are likely to prevent resonance capture or disrupt resonant chains after they form.

Here we build on \citet{macd18}'s proof of concept by investigating the establishment of resonant chains within systems arising from in-situ planet formation. We make use of and expand upon a suite of in situ formation simulations from \citet{macd20} that begin with planetary embryos in a depleted gas disk. \citet{macd20} identified a range of disk conditions that can account for the observed orbital and compositional properties of Kepler super-Earths and mini-Neptunes. We will investigate whether those same conditions produce resonant chains. In Section \ref{sec:plsysgen}, we summarize how we generated our model planetary systems that arose from in-situ planet formation. In Section \ref{sec:sysres}, we assess resonant chain outcomes. We discuss conditions needed to establish resonant chains in Section \ref{sec:conditions} and the statistics and observability of chains from our simulations in Section \ref{sec:stat}. We summarize our findings in Section \ref{sec:concl}.

\section{Generating Planetary Systems via In-Situ Formation}\label{sec:plsysgen}

We use and supplement the simulations of in situ formation from \citet{daws16} and \citet{macd20}. To summarize, we assume planetary embryos have grown from a reservoir of solid material transported from the outer disk into the region of formation. Our simulations begin as the gas disk starts to dissipate and embryos begin to interact gravitationally; solid material is no longer being brought in at significant levels from the outer disk.

We perform N-body integrations of planetary embryos lasting at 28 Myr using the mercury6 hybrid symplectic integrator \citep{Chambers:1996} in which the first 1 Myr includes eccentricity damping to mimic the dissipative effects of a depleted gaseous protoplanetary disk. The planetary embryos begin with zero eccentricity and masses:
\begin{eqnarray}\label{eq:Memb}
    M_{\rm emb}=\nonumber\\0.16M_\oplus\left(\frac{\Sigma_{z,1}}{10 \rm{ g/cm^{-2}}}\right)^{3/2}\left(\frac{a}{ \rm{AU}}\right)^{3(2+\alpha)/2}\left(\frac{M_*}{M_\odot}\right)^{-1/2},\nonumber\\
\end{eqnarray}
where the radial distribution of solids $\Sigma_z=\Sigma_{z,1}(a/\rm{AU})^\alpha$, $a$ is the semi-major axis, $M_\star$ is the stellar mass, $M_\odot$ is the Sun's mass, and $M_\oplus$ is the Earth's mass. The masses correspond to an initial spacing $\Delta_0=3$ in Hill radii, where
\begin{equation}\label{eq:space}
    \Delta \equiv {R_H}(a_2 - a_1),
\end{equation}
\noindent and
\begin{equation}\label{eq:rh}
    R_H \equiv \frac{a_1+a_2}{2} \Big(\frac{M_{\rm emb,1}+M_{\rm emb,2}}{3M_{\star}}\Big)^{1/3}.
\end{equation} 
Initial inclinations are set to $0.1h/\sqrt{3}$, where $h = \Big(\frac{M_{\rm emb,1}+M_{\rm emb,2}}{3M_{\star}}\Big)^{1/3}$, and initial longitudes of periapse, longitudes of ascending node, and mean anomalies are randomized.

While the gas is present, it damps planetary eccentricities and inclinations according to $\dot{e}/e=-1/\tau$ and $\dot{i}/i=-2/\tau$, where the damping timescale in years is
 \[
     \tau=0.003d\Bigg(\frac{a}{\rm au}\Bigg)^2\Bigg(\frac{M_{\odot}}{M_p}\Bigg) \times   
     \begin{cases} 
      1 & v\leq c_s, \\
      (v/c_s)^3 & v> c_s, i < c_s/v_K,\\
      (v/c_s)^4 & i > c_s/v_k,
   \end{cases}
 \]
\noindent where n is the planet's mean motion, the random epicyclic velocity $v = \sqrt{e^2+i^2} na$,  $c_s~=~1.29$km/s~($a$/au)$^{-1/4}$ is the gas sound speed \citep{Papaloizou2000,Kominami2002,Ford2007,Rein2012}, and $d$ is the depletion factor relative to the minimum mass solar nebula. The gas disk can also cause migration but we have confirmed through trial simulations that include migration that the slow migration expected in a depleted gas disk has a negligible effect on the final planet properties. To clarify the role of eccentricity damping, we use a $\dot{e}$ and $\dot{i}$ only following \citealt{wolf12}; the user-defined implemented forces in {\tt mercury6} have no direct effect on $a$.

The scenario we envision is that $1/d$ declines gradually until it reaches some threshold value, and the gas disk rapidly disappears (the photoevaporative switch model, e.g., \citealt{owen11,owen12}).  Here we approximate the dissipation process as a step function: we begin with $d$ at its threshold value for a 1 Myr, and subsequently $d=0$. The 1 Myr timescale represents the dissipation timescale at the end of the disk lifetime. We confirm that the step function approximation does not introduce sudden, spurious capture into or escape from resonance; in our simulated systems, capture occurs well before or after the damping force shuts off. The eccentricity damping acceleration is small, of order $10^{-6}$ the Keplerian acceleration.

The simulations are grouped into ensembles by degree of damping and summarized in Table 1. We list the number of systems with trios with integer period ratios within 2\% of 2:1, 3:2, 4:3, or 5:4 and number of systems with librating chains. We find that Ensemble Ed2 produces far more resonant chains than the others (for reasons we will explore in Section \ref{sec:conditions}); Ed2 is also the ensemble that we found can provide a good match to Kepler planets' observed orbital and compositional properties \citep{macd20}. We identify a range of $\Sigma_{z,1}$ (55--148 gcm$^{-2}$) within the ensemble that produces the majority of resonant chain systems and perform additional simulations (Ensemble Ed2+).

\begin{table*}[!ht]
\centering
\begin{tabular}{cccccc}
\hline
Name & Damping  & $\Sigma_{z,1}$ (gcm$^{-2}$) & \# of Simulations & Integer period ratio trios\footnote{Systems contain at least one trio where each pair is within 2\% of a 2:1, 3:2, 4:3, or 5:4 orbital period ratio.} & Librating chains\footnote{Systems contain at least one trio where each pair's two-body angle and/or the three-body angle is librating.}  \\ \hline
Ed4  & d=$10^4$ & 38--105         & 80                                     &0  & 0                             \\
Ed3  & d=$10^3$ & 38--105         & 80                                       & 2 &0                              \\
Ed2  & d=$10^2$ & 14--284         & 290                                      & 28 &8                             \\
Ed2+  & d=$10^2$ & 55--148         & 80                                       &      13                            & 6                             \\
Ed1  & d=10     & 14--284         & 240                                       & 14 &2                            \\
Ed0  & d=1      & 38--105         & 80                                       &0 &0                              \\ \hline
\end{tabular}
\caption{Suites of simulations in which we assessed the occurrence of planets in resonant chains. All surface density profiles have a power law slope of -1.5.}
\end{table*}

\section{Resonant Chain Outcomes of Oligarchic Growth}\label{sec:sysres}

We identify resonant chains in the formation simulations described in Section \ref{sec:plsysgen}. In Section \ref{subsec:types}, we describe the two types of resonant chains we look for and their associated resonant angles. In Section \ref{subsec:resresults}, we describe the process and results of identifying resonant chains in our simulations.

\subsection{Classifying resonant chains and angles}
\label{subsec:types}
We consider two types of resonant chains. The first type of chain is a set of at least three planets in which each successive pair of planets has at least one of its corresponding 2-body angles librating. For first order resonances, there are two possible resonant angles, and they consist of
\begin{eqnarray}
    \phi_{p+1:p,1,i}(1,2)=(p+1)\lambda_2-p\lambda_1-\varpi_1 \nonumber \\
        \phi_{p+1:p,1,o}(1,2)=(p+1)\lambda_2-p\lambda_1-\varpi_2 \nonumber \\
\end{eqnarray}
where $p$ is an integer corresponding to a $p+1:p$ resonance, $\lambda$ is the orbit mean longitude, $\varpi$ is the longitude of periapse, and the subscripts 1 and 2 refer to the inner and outer planet, respectively. In many cases, the planets' period ratios are near $p+1:p$, but if the longitude of periapse quickly precesses, the period ratio can be significantly different (e.g., \citealt{lithwick12}).

The second type of resonant chain is a set of three planets with a librating 3-body angle. Here we focus on 3-body angles that depend only on the planets' mean longitudes because this type of 3-body angle has been investigated for observed systems (e.g., \citealt{macd16}). These angles consist of consist of
\begin{equation}
\label{eqn:3body}
    \phi_{3b/p,q}(3,2,1)=p \lambda_3 -(p+q)\lambda_2 + q \lambda_1
\end{equation}
where the subscripts 1, 2, and 3 refer to the inner, middle, and outer planet, respectively. For commensurate period ratios:
\begin{equation}
    \frac{p}{q} = \frac{P_2/P_1 - 1}{1-P_2/P_3}
\end{equation}
Table \ref{tab:3body} contains a non-exhaustive list of three-body angles.


\begin{table}[!ht]
\centering
\begin{tabular}{l|l|l}
$P_2:P_1$ & $P_3:P_2$& Three-body \\
\hline
5:4 & 4:3 & $\phi_{3b/4,4}(3,2,1)$ \\
5:4 & 3:2 & $\phi_{3b/3,4}(3,2,1)$ \\
4:3 & 5:3 & $\phi_{3b/5,6}(3,2,1)$ \\
4:3 & 3:2 & $\phi_{3b/1,1}(3,2,1)$ \\
4:3 & 4:3 & $\phi_{3b/4,3}(3,2,1)$ \\
4:3 & 5:4 & $\phi_{3b/5,3}(3,2,1)$ \\
4:3 & 6:5 & $\phi_{3b/2,1}(3,2,1)$ \\
3:2 & 2:1 & $\phi_{3b/1,1}(3,2,1)$ \\
3:2 & 3:2 & $\phi_{3b/3,2}(3,2,1)$ \\
3:2 & 4:3 & $\phi_{3b/2,1}(3,2,1)$ \\
3:2 & 5:4 & $\phi_{3b/5,2}(3,2,1)$ \\
5:3 & 5:3 & $\phi_{3b/5,3}(3,2,1)$ \\
5:3 & 3:2 & $\phi_{3b/2,1}(3,2,1)$ \\
2:1 & 2:1 & $\phi_{3b/2,1}(3,2,1)$ \\
2:1 & 3:2 & $\phi_{3b/3,1}(3,2,1)$ \\
\end{tabular}
\caption{three-body angles (Eqn. \ref{eqn:3body}).\label{tab:3body}}
\end{table}

A triplet of planets can be in both types of resonant chain, the first type only, or the second type only (e.g., \citealt{char18}). The Galilean satellites (Io, Ganymede, and Europa) and GJ 876 are examples of systems with both librating two-body and three-body angles (e.g., \citealt{nels16}). For observed systems, it can be easier to determine if the three-body angle is librating than two-body angles because $\varpi$ can be challenging to measure. For example, the TRAPPIST-1 is known to have multiple librating three-body angles but it is unknown whether any of the two-body angles are librating \citep{luge17}.

\subsection{Resonant chains formed in simulations}
\label{subsec:resresults}

\begin{figure}[ht!]
\plotone{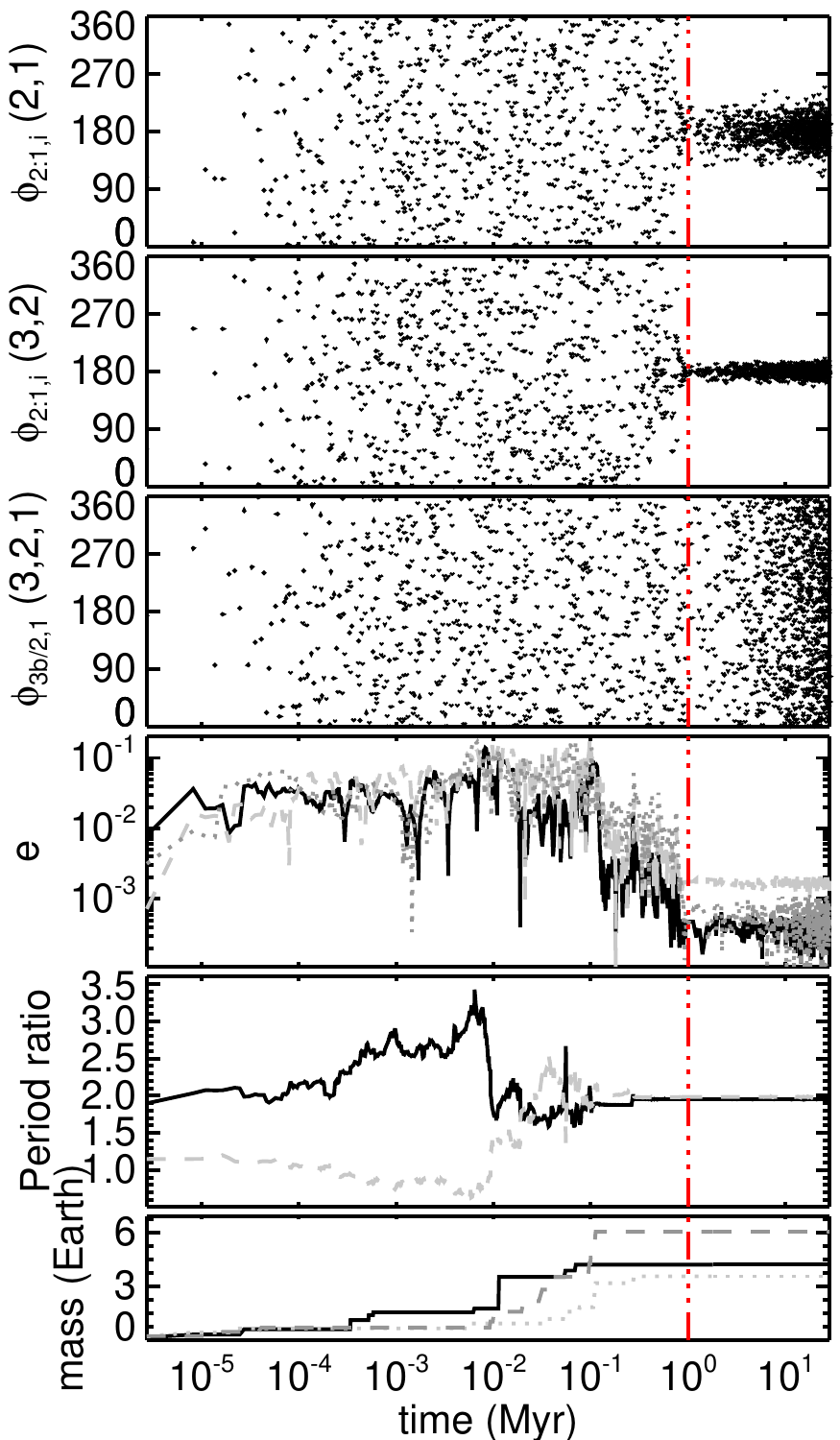}
\caption{Example of resonant chain formed in situ with interlocking two-body angles (row 1 and 2) but no librating three-body angle (row 3). The vertical dot dashed line marks the end of the gas disk stage. During the gas disk stage, planets grow through mergers (bottom row) and excite each others' eccentricities (row 4). Once they grow sufficiently isolated, their eccentricities damp and they capture into two-body mean motion resonance 0.2 Myr before the dissipation of the gas disk. The resonant angle of the inner pair involves the periapse of the inner planet, not the middle planet (i.e., $\theta_{2:1,o}(2,1)$, not shown, does not librate). Note that the final system contains 12 planets within 1 AU, six interior and three exterior to the resonant triplet.}
\label{fig:2bodonly}
\end{figure}

\begin{figure}[ht!]

\plotone{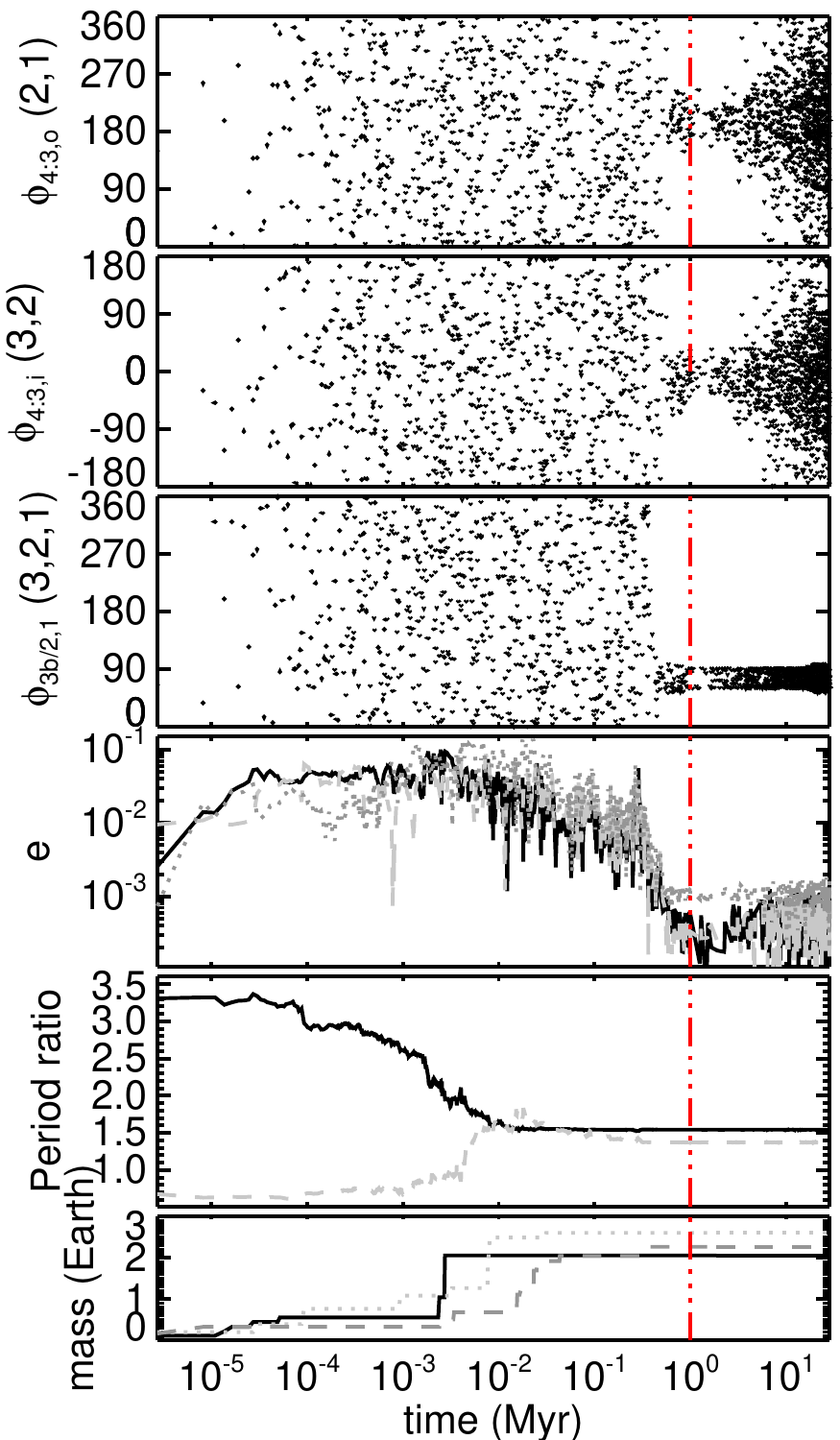}
\caption{Example of resonant chain formed in situ with librating three-body angle (row 3) but without libration of the corresponding two-body angles (row 1 and 2). The vertical dot dashed line marks the end of the gas disk stage. During the gas disk stage, planets grow through mergers (bottom row) and excite each others' eccentricities (row 4). Once they grow sufficiently isolated, their eccentricities damp and they capture into resonance. After the gas disk stage, eccentricities slowly grow due to perturbations from other planets in the system and the two-body libration amplitudes grow until the planets are no longer in resonance; the three-body angle remains tightly librating. Note that the final system contains 14 planets within 1 AU, five interior and six exterior to the resonant triplet.}
\label{fig:3bodonly}
\end{figure}

\begin{figure}[ht!]

\plotone{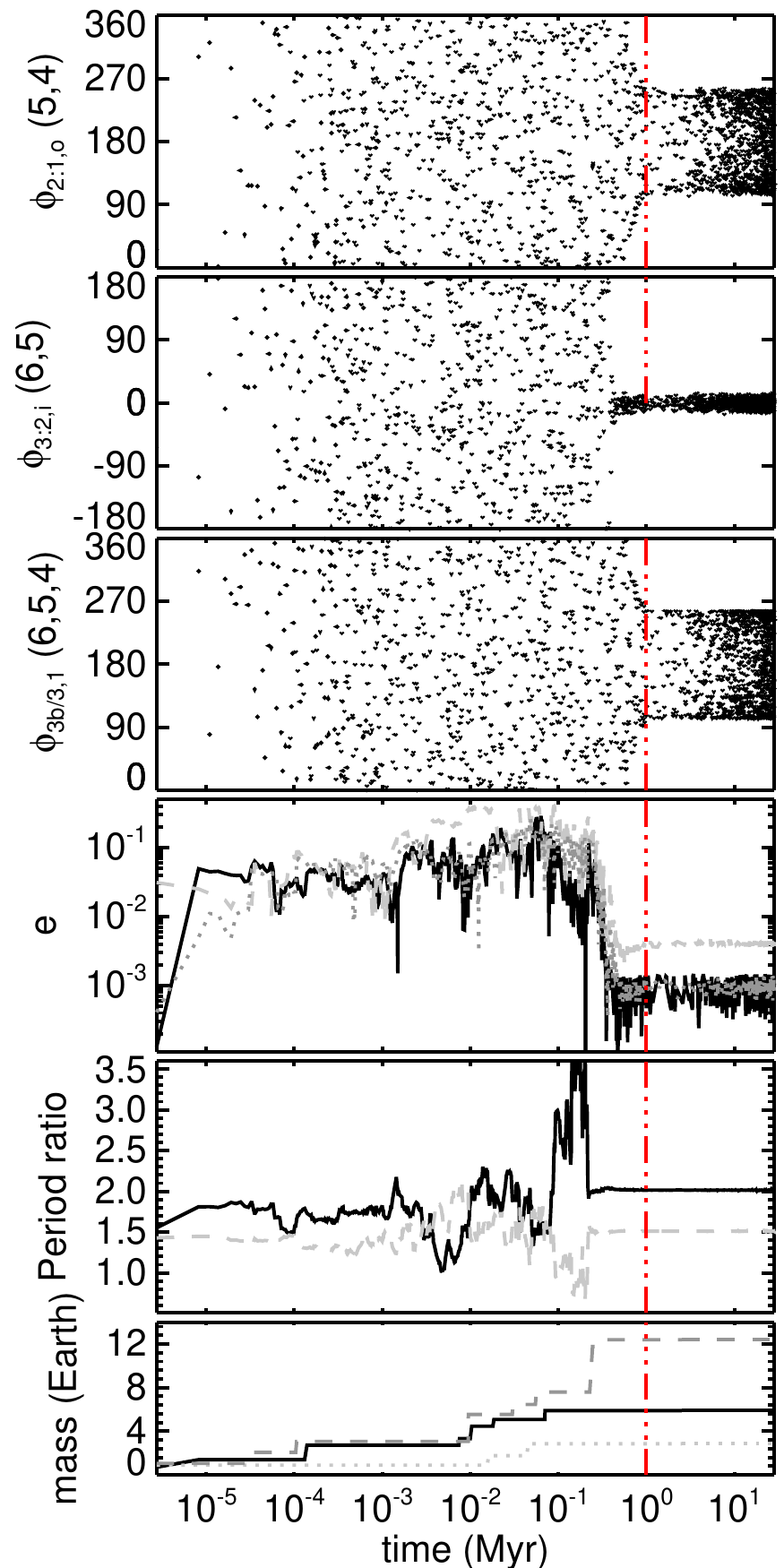}
\caption{Example of resonant chain formed in situ with librating three-body angle (row 3) and also libration of the corresponding two-body angles (row 1 and 2). The vertical dot dashed line marks the end of the gas disk stage. During the gas disk stage, planets grow through mergers (bottom row) and excite each others' eccentricities (row 4). Once they grow sufficiently isolated, their eccentricities damp and they capture into resonance 0.4 Myr before the dissipation of the gas disk}. Note that the final system contains 7 planets within 1 AU, four interior to the resonant triplet shown here. Six of the planets are involved in a resonant chain.
\label{fig:both}
\end{figure}

To systematically identify resonant chains, we generate plots of two-body and three-body angles and assess each by eye for libration. We list the resonant chains in Table \ref{tab:chains} and show examples in Figures \ref{fig:2bodonly}, \ref{fig:3bodonly}, and \ref{fig:both}. We identify triplets in both types of resonant chains. Most chains contain only three planets, but several contain four or more, including one chain of five planets and two chains of six planets, reminiscent of the TRAPPIST-1 system. We see resonant chains across the full range of simulated orbital periods. 

We find that simulated systems with both types of resonant chains (i.e., librating three-body angles and successive pairs of librating two-body angles) are relatively uncommon: most resonant chains are either one type or the other. In order for a librating pair of two-body resonances to dictate the libration of the three-body angle, $\phi_{p+1:p,1,o}(1,2)$ and $\phi_{p+1:p,1,i}(2,3)$ must both librate. For most of our simulated chains, $\phi_{p+1:p,1,o}(1,2)$ does not librate. One of our six planet chains is an exception: the two-body angles $\phi_{2:1,o}(5,4)$ and $\phi_{3:2,i}(6,5)$ both librate, so three-body angle $\phi_{3b/3,1}(6,5,4)$ librates as well (Fig. \ref{fig:both}).

Chains of interlocking two-body resonances include the 2:1, 3:2, 4:3, 5:4, and 6:5. An example is shown in Fig. \ref{fig:2bodonly}. The two pairs can have similar period ratios (e.g., both pairs of planets in the 2:1 MMR, like the observed system GJ 876) or different period ratios (e.g., one pair in the 4:3 and one in the 3:2, like the observed system Kepler-223). Period ratios can either be narrow or wide of commensurability. For example, one system features two pairs of 3:2 MMR with period ratios of 1.48 and 1.47 (narrow of commensurability), and another features a 4:3 pair with a period ratio of 1.34 and 5:4 pair with a period ratio of 1.27 (both wide of commensurability). As expected from stability, planets with period ratios closer to 1 tend to be less massive. For  example, a 3.3 $M_\oplus$ and 0.9  $M_\oplus$ pair exhibits libration of the 5:4 two-body angle, and a 8.6 $M_\oplus$ and 9.3 $M_\oplus$ pair exhibits libration of the 2:1.

We find a number of systems with librating three-body angles where the corresponding two-body angles do not librate. For example, a system with an inner pair with period ratio near 4:3 and outer pair with period ratio near 6:5 exhibits libration of $\phi_{3b/2,1}(3,2,1)$ but not $\phi_{4:3,i/o}(2,1)$ or $\phi_{6:5,i/o}(2,1)$. We find only a couple systems (including one of the six planet chains) with libration of both the three-body angle and the corresponding two-body angles. 

In summary, resonant chains of both types naturally can arise from our simulations of in situ formation when planets happen to form near consecutive integer period ratios. In the next section, we will explore why some planets near integer period ratios do not end up in resonant chains and which formation conditions are most favorable for capture.

\section{Conditions favorable for producing MMR chains}
\label{sec:conditions}

All of our simulations of in situ formation that produced resonant chains met the following conditions: 1) planets must achieve a chain of near-integer period ratios during the gas disk stage, 2) the planets must experience eccentricity damping that captures them into resonance, 3) the planets must stay in resonance as the gas disk dissipates, and 4) the system must avoid subsequent giant impacts, eccentricity excitation, and chaotic diffusion that would dislodge the planets from resonance. 

\begin{figure}
\includegraphics[width=3.5in]{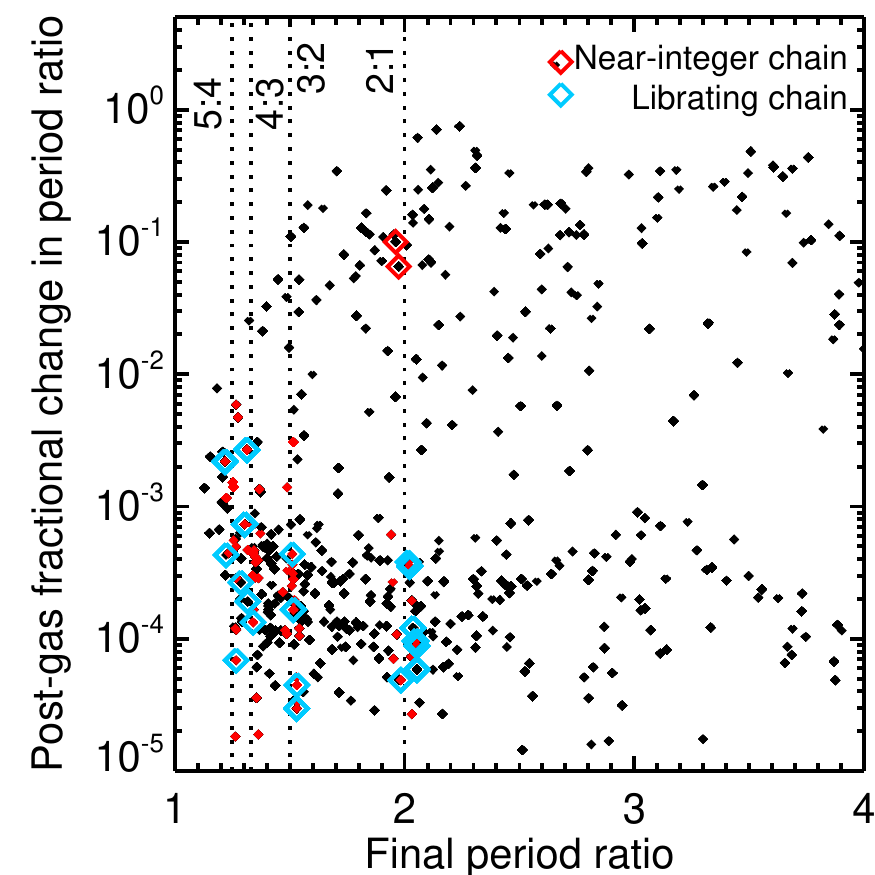}
\caption{Fractional change in period ratio during the post-gas stage vs. final period ratio from ensemble Ed2+. Planets in chains of two or more pairs near integer period ratios (but not necessarily with librating two-body angles) are marked in red; those with librating resonant angles are outlined in blue. Larger red symbols indicate planets in simulation that achieve their near integer period ratios after the gas disk stage. All other chains of near period ratios are established during the gas disk stage.}
\label{fig:ratchange}
\end{figure}

\begin{figure}[ht!]
\plotone{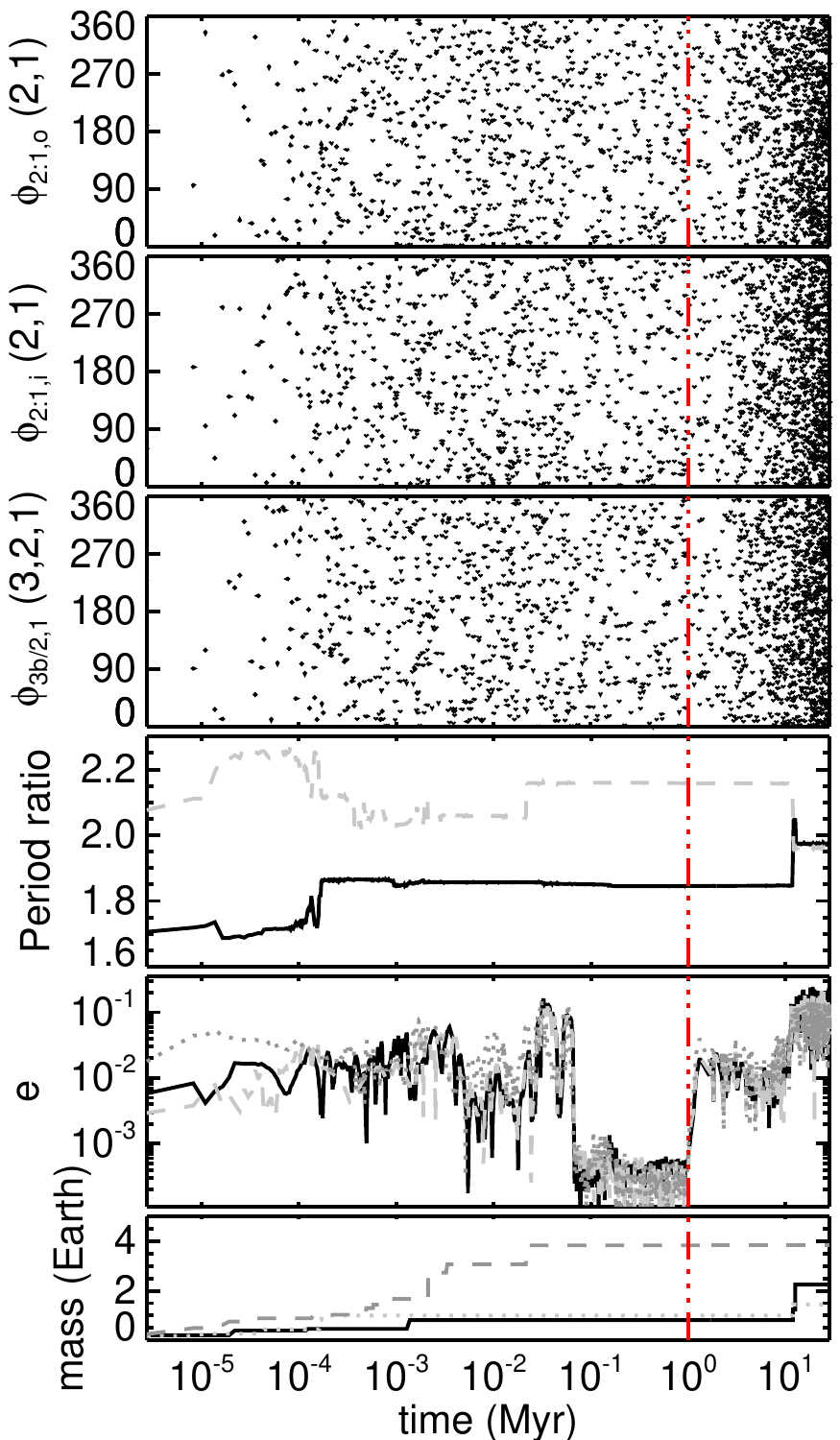}
\caption{Example of system where planets do not achieve a chain of near-integer period ratios until after the gas disk stage, when late collisions alter their orbital periods. With no dissipation, they are not captured into resonance.}
\label{fig:prat_late}
\end{figure}

Planetary systems fail the first criterion when they only achieve a chain of near-integer period ratios after the gas disk stage. In Fig. \ref{fig:ratchange}, we plot the fractional change in period ratio during each simulated system's post-gas stage evolution from ensemble Ed2+. For this ensemble, planets in chains of near-integer period ratios (defined as a period ratio within 3\% of an integer value) tend to reach those period ratios during the gas-disk stage. We only find one exception, shown in Figure \ref{fig:prat_late}. In this system, a planetary trio continue to undergo mergers and reach their final masses and period ratios at around 10 Myr after the gas disk dissipates. Without dissipation, they do not capture into resonance. 

More generally, under ensemble Ed2+'s gas-disk conditions, most growth occurs during the gas-disk stage. The average Hill spacing is 14 at the end of the gas-disk stage and only increases, due to mergers among a small subset of planets, to 19 after 27 Myr of post-gas evolution. Of course, under other disk conditions (i.e., leading to planets undergoing most of their growth after the gas disk stage), chains of near-integer period ratios may be primarily established after the gas disk stage; but such conditions cannot account for other observed properties of \kep systems \citep{macd20}.

\begin{figure}[ht!]
\plotone{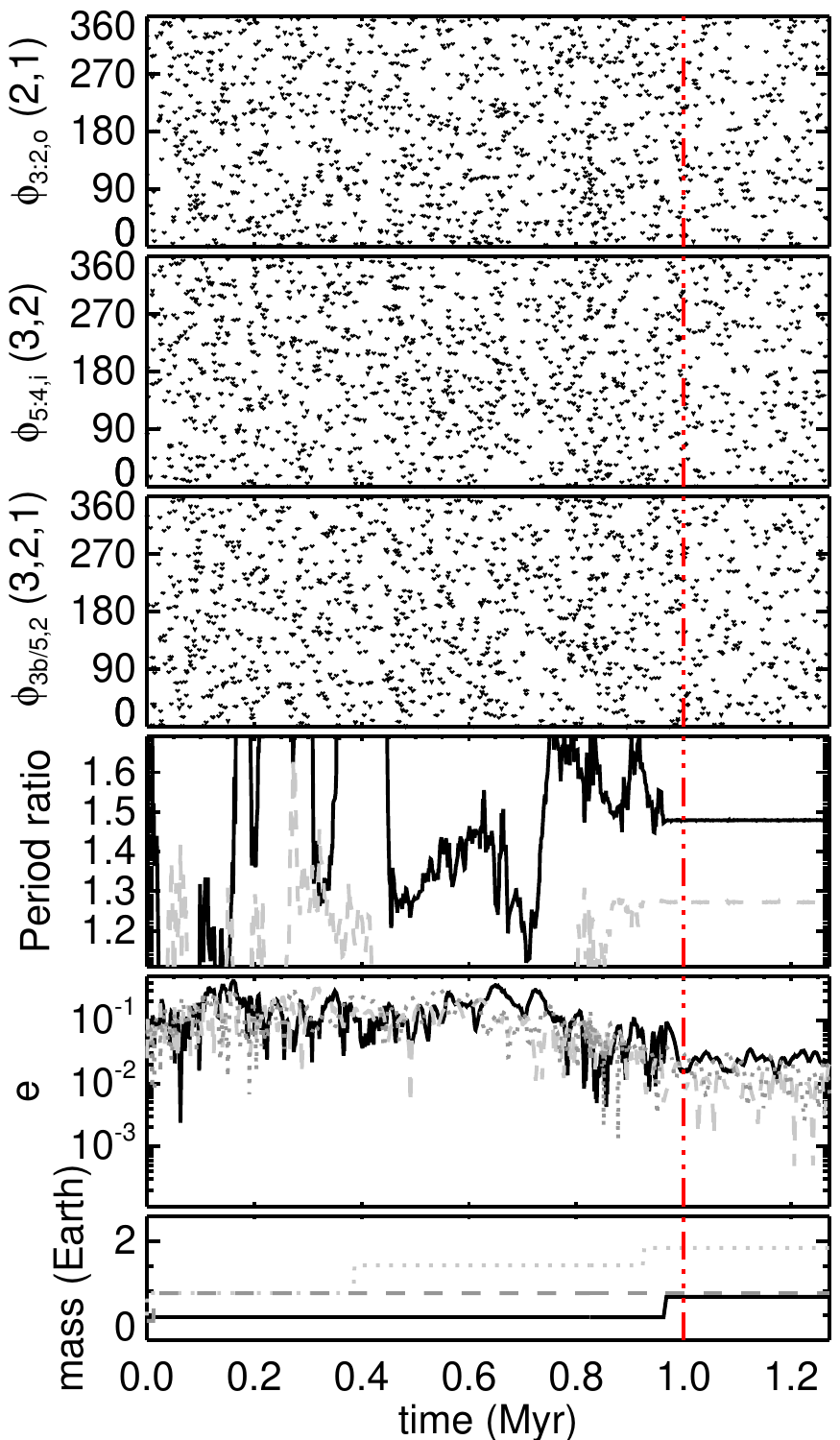}
\caption{Example of system where planets do not achieve a chain of near-integer period ratios until just before the end of the gas disk stage. There is insufficient time for eccentricity damping and resonance capture before the gas disk dissipates.}
\label{fig:end}
\end{figure}

\begin{figure}[ht!]
\plotone{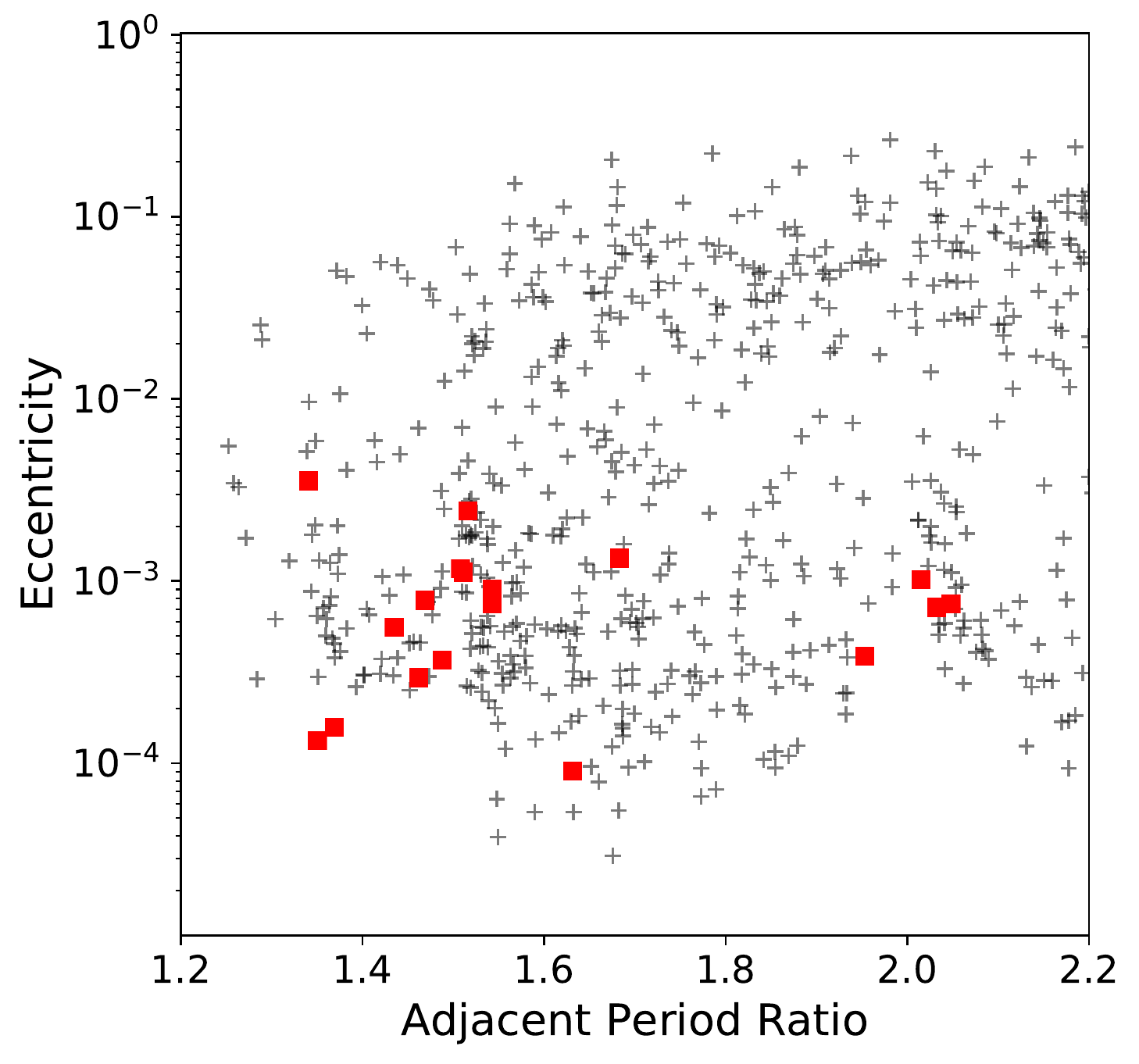}
\caption{Planets participating in resonant chains (red squares) have distinctly lower eccentricities than the overall resulting distribution of planet eccentricities following formation in depleted gas disk conditions and subsequent dynamical evolution (ensemble Ed2).}
\label{fig:prvsec}
\end{figure}

\begin{figure}[ht!]
\plotone{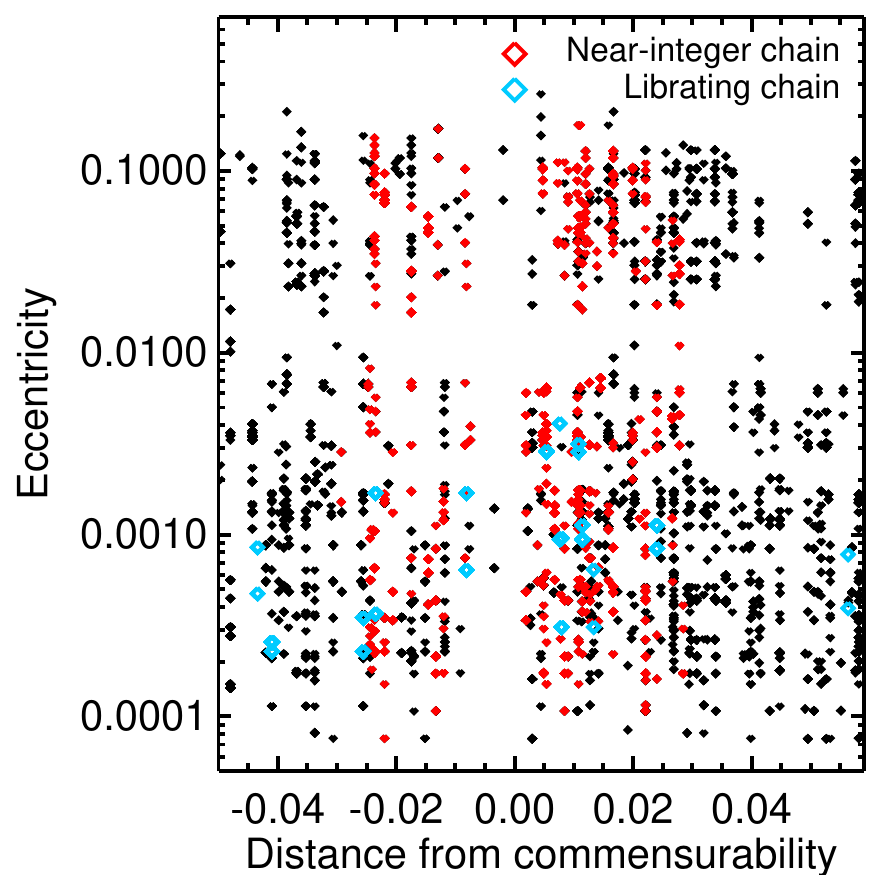}
\caption{Librating resonant chains (blue) have lower eccentricities than chains with near commensurabilities (red) (Ensemble Ed2+). Distance from commensurability is calculated as $\frac{\rm Period_{\rm outer}/Period_{\rm inner}}{1+1/p} -1$ for the nearest $p+1:p$ first order resonance, where the period ratio is averaged over 0.03 Myr.
\label{fig:edelta}}
\end{figure}

Sometimes planets reach near-integer period ratios during the gas disk stage but do not experience sufficient eccentricity damping for resonance capture. Eccentricity damping is necessary because our in simulations of in situ formation do not include migration, so damping to low eccentricities allows for libration at a wider range of period ratios (e.g., \citealt{del12}). Fig. \ref{fig:end} shows an example of a trio where two of the planets experience their last mergers right before the end of the gas disk stage at 1 myr. Their eccentricities are damped slightly in the short remaining time, but not enough for resonance capture. The need for eccentricity damping for resonance capture likely contributes to a trend in our simulations:  planets captured into resonant chains in our simulations end up with lower eccentricities at a given period ratio (Fig. \ref{fig:prvsec}). Figure \ref{fig:edelta} zooms in on the period ratios near first order commensurabilities for simulated planets. Eccentricity damping enables libration away from exact commensurability and causes a gap in the period ratio distribution just inside commensurability (e.g., \citealt{lith12b}). We do not see an obvious trend between smaller eccentricities and wider distances from commensurability, but the diversity of simulated planet masses could obscure such a trend.

\begin{figure}[ht!]
\plotone{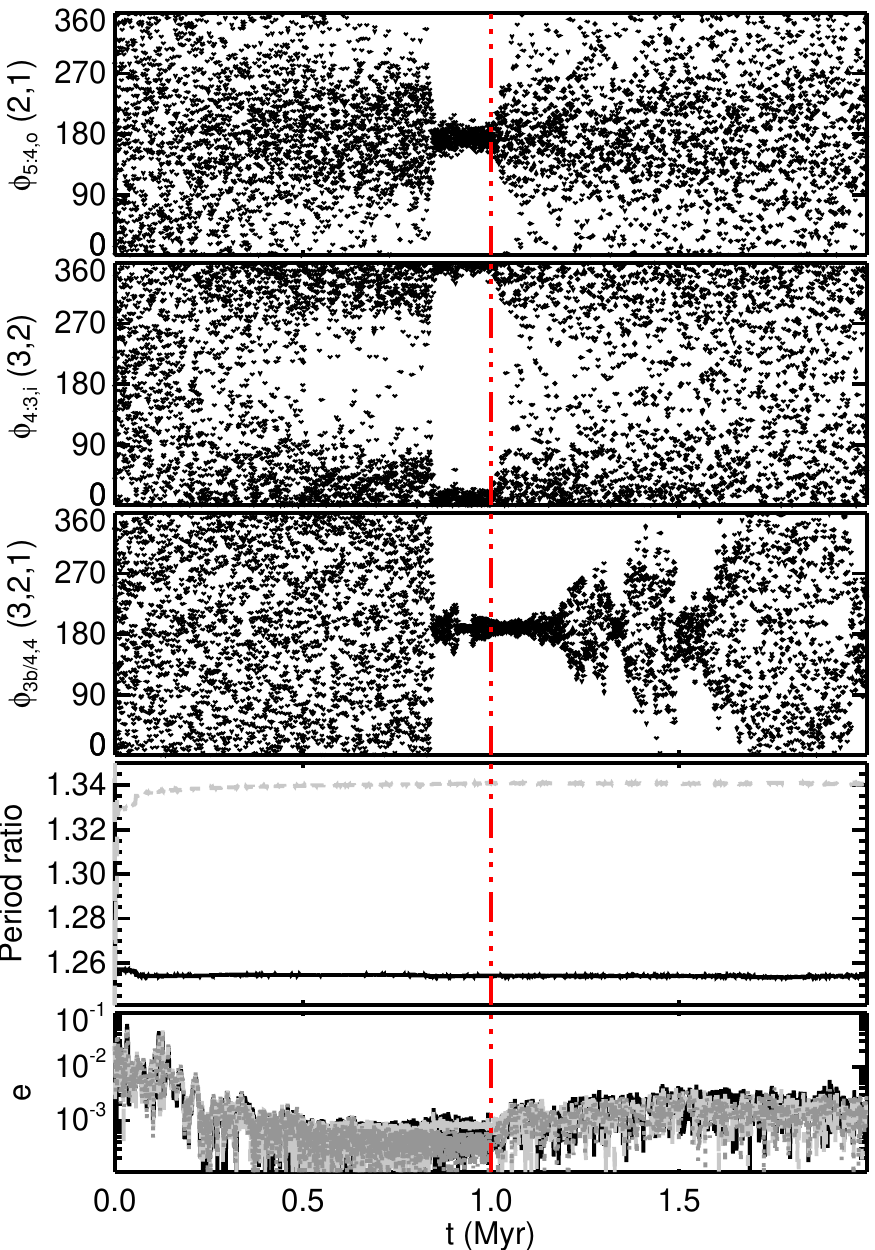}
\caption{Example of system that escapes resonance shortly (0.05 Myr) after the gas disk dissipates (red dashed line).}
\label{fig:crit3}
\end{figure}

Some resonant chains lose their libration when the gas disk dissipates. We show an example in Fig. \ref{fig:crit3}. During the gas disk stage, the two body and three body angles achieve a low amplitude libration. Although nothing dramatic like a collision happens within the system when the gas disappears -- the period ratios remain constant (fourth panel) -- the two-body resonant angles cease to librate shortly after (panels 1--2) and the eccentricities grow modestly (panel 5). After about 0.5 Myr, the three body angle circulates as well (panel 3). The trio may get dislodged from resonance after the gas disk stage because it is no longer cushioned by the gas from perturbations of other planets in the system. 

Other systems lose their libration on a longer timescale. We find that two-body angles can be dislodged from resonance while keeping the corresponding three-body angle intact. Among our simulations, we find several cases where trios escape two-body resonances but remain in their three-body resonance. Fig. \ref{fig:3bodonly} shows an example where the two-body angles gradually escape resonance but the three-body angle remains tightly librating. In another resonant chain, the two-body angles are dislodged when a pair of interior planets uninvolved in the chain undergo a collision -- leading to a spike in the resonant planets' eccentricities -- but the three-body angle remains tightly librating.

To capture and maintain planets in resonant chains, disk conditions must allow planets to reach their final masses and experience eccentricity damping during the gas disk stage, with few disturbances afterwards. These are the same conditions that enable tightly packed planets with period ratios less than 2 \citep{macd20}, favorable for achieving integer period ratios. In the next section, we will quantify just how commonly resonant chains are produced under the right conditions.

\section{Statistics and observability}
\label{sec:stat}

Our simulations of in situ formation establish resonant chains under the right disk conditions. Here we assess how often we expect those chains to be observable in the \kep sample. We follow \citet{macd20} to assess the observability of each system, forward modeling both detection efficiency among \kep targets  \citep{Burke2015,Christiansen2015,chri16} using parameters for the DR25 \kep catalog \citep{thom18}, the  \citet{Weiss2014} mass-radius relationship, and geometric selection effects for $10^6$ randomly oriented copies of each simulated system. Each system has of order 20,000 realizations with one or more detected transiting planets. When comparing the observed sample of \kep targets, we use the 59,356 stars with stellar effective temperature $4100K < T_{eff}<6100K$, stellar $logg >4$, and Kepler magnitude $<$ 15.

For each simulated resonant chain, in Table \ref{tab:chains} we report $N_{\rm Kep}$, the number of detections of the entire resonant chain multiplied by the number of \kep targets meeting our selection criteria and divided by $10^6$. We can interpret $N_{\rm Kep}$ as the expected number of \kep detections (i.e., where each planet in the chain transits and has sufficient signal to noise to be observable around that star) if every \kep target meeting our selection criteria had an exact but randomly orientated copy of this planetary system. 

We also compute a related quantity, $f_{\rm ch,1+}$, the number of detections of the entire resonant chain divided by the total number of systems with one or more transiting planets generated by the ensemble, and $f_{\rm ch,3+}$, the number of detections of the entire resonant chain divided by the total number of systems with three or more transiting planets generated by the ensemble. 

As expected, resonant chains of larger planets at shorter orbital periods are more commonly ``detected." For example, we do not expect to ever detect the resonant chain where the outer planet has a mass of $0.5 M_\oplus$ and period of 316 days. In contrast, if every \kep star had the resonant chain formed in our simulations of six short period super-Earths (periods between 3--25 days and masses between 1.5--4 $M_\oplus$), we would expect to detect $N_{\rm Kep}$ = 247.

The quantities $f_{\rm ch,1+}$ and $f_{\rm ch,3+}$ give us a sense for whether our simulations are producing enough resonant chains to be consistent with observations. Consider Ensemble Ed2+, which contains 80 simulations, six of which produce librating resonant chains. Of the simulated transiting systems containing one or more planets, 0.6\% have transiting, librating resonant chains; of the simulated transiting systems containing three or more planets, 3\% have transiting, librating resonant chains. The \kep candidate catalog with our selection criteria applied contains 1110 systems with one or more transiting planets and 109 systems with three or more transiting planets. Ensemble Ed2+ contains a subset of the disk conditions employed by \citet{macd20} to reproduce the Kepler sample. That subset produces about 44\% of simulated systems with one or more detected transiting planets and 71\% of those with three or more detected transiting planets. Therefore we expect of order 3 observed systems with librating resonant chains. We know of at least two observed \kep super-Earth systems with librating resonant chains (Kepler-223, \citealt{mill16}, and Kepler-80, \citealt{macd16}; MacDonald et al., in prep), and there are other possible chains (based on period ratio) for which the libration status is unknown\footnote{The libration status is typically determined by performing longer term integrations of a random subset of solutions from the fit to light curves and/or mid-transit times. The libration status is unknown when some solutions librate but others circulate. This uncertainty in status can arise even when the instantaneous value of the resonant angle is well measured, if other properties like mass and eccentricity are uncertain.}. The in situ simulations therefore produce librating resonant chains at a rate consistent with those detected by the \kep Mission, but would be underproducing librating resonant chains if many other \kep trios have undetected libration.

The \kep DR25 candidate catalog with our selection criteria applied contains 10 trios -- among the 110 systems with three or more transiting planets -- where each pair is within 2\% of a 2:1, 3:2, 4:3, or 5:4 orbital period ratio. Similarly, the Ed2+ ensemble contains 784 such trios out of 8161 systems with three or more transiting planets. Therefore our simulations are producing chains of integer period ratios in line with what is observed.

\section{Conclusions}\label{sec:concl}

Here we demonstrated that resonant chains can naturally arise in simulations of oligarchic growth in a depleted gas disk. Resonant chains occur when three nearby planets achieve a chain of near-integer period ratios during the gas disk stage, experience eccentricity damping that captures them into resonance, stay in resonance as the gas disk dissipates, and avoid subsequent giant impacts, eccentricity excitation, and chaotic diffusion that would dislodge the planets from resonance. Some chains contain pairs of librating two-body angles, some contain librating three-body angles, and some contain both. Planets in chains of two-body resonances are captured into resonances as tight as 5:4 and as wide as 2:1. Generally the three-body angles are most robust against being dislodged from resonance.

The types of disk conditions that form tight period ratios \citep{macd20} are also the ones that can establish resonant chains. Near the end of the gas disk stage, oligarchs are no longer completely gravitationally cushioned from each other by the gas disk and can grow through mergers. They finish forming during the gas stage and can accrete low mass gas envelopes. They remain on their compact, coplanar, low eccentricity orbits after the gas disk dissipates, throughout the stellar lifetime.  \citet{macd20} found in their simulations that adjacent planets with period ratios less than 2 tend to have larger radii than planets in less compact configurations. The planets captured into resonant chains are part of this population, and therefore we expect them to also have larger radii at a given mass (lower bulk density) than the general population. Therefore if future observations identify differences in composition between resonant and non-resonant planets, it is possible that those differences could be a result of different disk conditions for in situ formation (e.g., \citealt{macd20}) rather than formation outside vs. inside the ice line. 

The resonant chains produced in our simulations generally reside in systems with other planets that are not part of the chain and that may or may not also transit. The high intrinsic multiplicity -- ranging from 5 to 14 planets interior to 1 AU -- results from our assumed initially continuous distribution of solids. To determine whether high underlying multiplicity is a testable prediction of in situ formation, future studies must explore whether initial distributions with gaps or rings can also produce resonant chains while matching other Kepler observables like transit multiplicity.

The properties of resonant chains that emerge from our simulations are consistent with our current knowledge of observed systems. They exhibit a range of libration amplitudes, from narrow (e.g., the three-body angle in Fig. \ref{fig:3bodonly}) to wide (e.g., the two-body angles in Fig. \ref{fig:2bodonly}, which eventually escape from resonance). In situ formation apparently does not preclude the tight libration of three-body angles observed in some real systems (i.e., \citealt{mill16,macd16}). True resonant chains and trios of near-integer period ratios are prevalent enough in our simulations to account for those observed (Section \ref{sec:stat}). However, if many more observed trios of integer period ratios are found to be librating, our simulations would be underproducing resonant chains. We recommend future investigations of whether to expect systematic differences in the properties of resonant chains expected from in situ formation vs. long distance migration.

\citet{macd18} concluded from their case studies of four observed systems that from the presence of a resonant chain alone, we cannot deduce whether the planets formed in situ and were captured through short distance migration or eccentricity damping only or whether they formed much further out and migrated in. Here we generalize that conclusion by identifying resonant chains formed in situ in our suite of simulations of in situ formation. Although disturbances from other planets in the system succeed in dislodging chains or preventing capture in some systems, in other cases resonant chains manage to form and survive. 

\acknowledgments

We thank the referee, Dan Tamayo, for the helpful report. The authors were supported in part by NASA Exoplanet Research Program Grant No. 80NSSC18K0355 and the Center for Exoplanets and Habitable Worlds at the Pennsylvania State University. The Center for Exoplanets and Habitable Worlds is supported by the Pennsylvania State University, the Eberly College of Science, and the Pennsylvania Space Grant Consortium. SJM also acknowledges support from the Missouri State University. RID acknowledges support from the Alfred P. Sloan Foundation's Sloan Research Fellowship. MGM acknowledges that this material is based upon work supported by the National Science Foundation Graduate Research Fellowship Program under Grant No. DGE1255832. Any opinions, findings, and conclusions or recommendations expressed in this material are those of the author and do not necessarily reflect the views of the National Science Foundation. This research made use of the NASA Astrophysics Data System Bibliographic Services and computing facilities from Penn State's Institute for CyberScience Advanced CyberInfrastructure. 

\appendix
\section{Table of Resonant Chains}

Resonant chains produced in the simulations are listed in Table \ref{tab:chains}. 

\begin{longrotatetable}
\begin{deluxetable*}{llrrrrrrcccccllll}
\tablecaption{Simulated resonant chains\label{tab:chains}}
\tablehead{}
\startdata
Ens. & $\Sigma_{z,1}$& $P_1$  & $P_2$ & $P_3 $ &$P_4$&$P_5$&$P_6$&$\frac{P_2}{P_1}$&$\frac{P_3}{P_2}$&$\frac{P_4}{P_3}$&$\frac{P_5}{P_4}$&$\frac{P_6}{P_5}$&Librating angle(s) & $N_{\rm  Kep}$ & $f_{\rm ch,1+}$& $f_{\rm ch,3+}$\\
&&(d)&(d)&&&&&&&&&&&&$(10^{-4})$&$(10^{-3})$\\
&&$m_1$&$m_2$&$m_3$&$m_4$&$m_5$&$m_6$\\
&&$(M_\oplus)$\\
\hline\hline
Ed2&17& 65&   24.6&   33.1&  40.0& &              &       1.35&   1.21&   &      & &              $\phi_{3b/2,1}(3,2,1)$&    31  & 2 & 1  \\ 
&&0.7&2.1&1.6&&&&4:3&5:4\\
\hline
Ed2&   66&    28.3&  43.6&  59.9&  &       &       &       1.54&   1.37&   &       &       &       $\phi_{3b/2,1}(3,2,1)$&    33 & 2 & 1 \\
&&2.1&2.6&2.3&&&&3:2&4:3\\
\hline
Ed2+&   71  &  45.6&  89.1&  177&  &       &       &       1.95&   1.98&   &       &       &       $\phi_{2:1,i}(2,1),\phi_{2:1,i}(3,2) $& 58&2&1\\
&&4.2&3.6&6.1&&&&2:1&2:1\\
\hline
Ed2&     74&  17.7&  23.5&  32.8&  &       &       &       1.38&   1.34&   &       &       &       $\phi_{3b/3,2}(3,2,1)$&   14 & 0.9 & 0.6 \\
&&1.8&1.8&1.3&&&&4:3&4:3\\
\hline
Ed2+&   84  &  21.9&  29.4&  37.3&  &       &       &       1.34&   1.27&   &       &       &       $\phi_{4:3,i}(2,1),\phi_{5:4,o}(3,2)$&  11&0.4&0.2\\
&&0.9&3.3&0.9&&&&4:3&5:4\\
\hline
Ed2&   88&    12.1&  18.3&  28.0&  &       &       &       1.52&   1.53&   &  &&       $\phi_{3:2,i}(2,1),\phi_{3:2,o}(2,1)$ &  200    &   9    &7\\ 
&&      1.8 &   2.1    & 2.4      &       &       &       &         3:2     &  3:2 &    &       &       &  \\
\hline
Ed2&   92&    145&  213&  316&  &       &       &       1.47&   1.48&   &       &       &       $\phi_{3:2,i}(2,1),\phi_{3:2,o}(2,1)$ & 0&0&0\\ 
&&      3.0 &   5.0    & 0.5      &       &       &       &         3:2     &  3:2 &    &       &       &  $\phi_{3:2,o}(3,2)$\\
\hline
Ed2&   96&    17.3&  21.8&  33.1&  &       &       &       1.26&   1.53&   &   && $\phi_{3b/3,4}(3,2,1)$ &  63   &    3   &     2 \\ 
&&      1.2 &   2.6    & 2.0      &       &       &       &         5:4     &  3:2 &    &       &       &  \\
\hline
Ed2+&    109&   111&  148&  225&  &       &       &       1.34&   1.52&   &       &       &       $\phi_{3b/1,1}(3,2,1)$&    6&0.2&0.1 \\
&&2.6&4.7&5.8&&&&4:3&3:2\\
\hline
Ed2&    112&168&   254&  340&  &       &       &       1.51&   1.34&   &    && $\phi_{3:2,o}(2,1),\phi_{4:3,i}(3,2)$&   0.1   &     0.004  &      0.002 \\
&&6.3&2.2&1.4&&&&4:3&3:2&&&&$\phi_{3b/2,1}(3,2,1)$\\
\hline
Ed2+&   112&    3.23&   4.39&   7.19&   11.7&  17.1&  24.7&  1.36&   1.64&   1.63&   1.46&   1.44&   $\phi_{4:3,i}(2,1),\phi_{3:2,i}(5,4)$&247\tablenotemark{a} &8\tablenotemark{a} &5\tablenotemark{a} \\ 
&&   1.5&2.4    &   2.2    &     3.2  &  3.5     &     4.0  &   4:3    & 5:3      &  5:3     &  3:2     &3:2              &       $\phi_{3:2,i}(6,5),\phi_{3b/2,1}(5,4,3)$\\
\hline
Ed2+&   133&    11.9&  27.4&  56.2&  114&  229&  346&  2.31&   2.05&   2.02&   2.01&   1.51&   $\phi_{2:1,i}(2,1),\phi_{2:1,i}(3,2)$\\ 
&&    6.3   &    8.6   &9.3       &  5.9     &  2.9     &     12  &    2:1   & 2:1      & 2:1      & 2:1      &3:2       &       $\phi_{2:1,i}(4,3),\phi_{2:1,o}(5,4)$&8\tablenotemark{b}&0.3\tablenotemark{b}&0.2\tablenotemark{b}\\        
&&       &       &       &       &       &       &       &       &       &       &       &       $\phi_{3:2,i}(6,5),\phi_{3b/3,1}(6,5,4)$\\    
&&       &       &       &       &       &       &       &       &       &       &       &      $\phi_{2:1,i}(5,4),\phi_{3:2,o}(6,5)$\\
\hline
Ed2+&   146&    101&  144&  194&  265&  420&  &       1.43&   1.34&   1.37&   1.58&   &       $\phi_{3:2,i}(2,1),\phi_{4:3,o}(3,2)$&0&0&0\\
&&     2.7  &    11.5   &   1.2    &    4.8   &    3.2   &       &   3:2    &4:3       & 4:3      &3:2       &       &  $\phi_{3b/1,1}(5,4,3) $ \\
\hline
Ed2&    166&123&   169&  267&  &       &       &       1.37&   1.58&   &       &       &       $\phi_{4:3,o}(2,1),\phi_{3b/1,1}(3,2,1)$&   0.7 & 0.02&0.008 \\
&&14.7&2.4&9.8&&&&4:3&3:2&&&&\\
\hline
Ed1& 228&    19.7&  30.0&  40.7&  &       &       &       1.53&   1.36&   &       &       &       $\phi_{3b/2,1}(3,2,1)$&  75&2&1 \\
&&12&1.3&1.7&&&&3:2&4:3&&&&$\phi_{3:2,o}(2,1),\phi_{4:3,i}(3,2)$\\
\hline
Ed1&    230&   94.4&  129&  202&  &       &       &       1.36&   1.57&   &       &       &       $\phi_{3b/1,1}(3,2,1)$&  58&2&1 \\
&&11&4.1&15&&&&4:3&3:2\\
\enddata
\tablenotetext{a}{For at least one subset detected (3--5 or 4--6), $N_{\rm  Kep} =327$,  $f_{\rm ch,1+} = 11 * 10^{-4}$, and $f_{\rm ch,3+} = 7*10^{-4}$.}
\tablenotetext{b}{For at least one subset detected (1--3, 2--4, 3--5, or 4--6), $N_{\rm  Kep} =964$,  $f_{\rm ch,1+} = 44 * 10^{-4}$, and $f_{\rm ch,3+} = 20*10^{-4}$.}

\end{deluxetable*}
\end{longrotatetable}

\end{document}